\documentclass[longnamesfirst,apjl]{emulateapj}
\usepackage{apjfonts,natbib,epsfig}
\tighten

\newcommand{\beq}{\begin{equation}}
\newcommand{\eeq}{\end{equation}}
\newcommand{\bea}{\begin{eqnarray}}
\newcommand{\eea}{\end{eqnarray}}

\begin{document}
\title{Weibel Filament Decay and Thermalization in Collisionless Shocks \\ and Gamma-Ray Burst Afterglows}
\author{Milo\v s Milosavljevi\'c and Ehud Nakar}
\affil{Theoretical Astrophysics, Mail Code 130-33, California Institute of Technology, 1200 East California Boulevard, Pasadena, CA 91125  }
\righthead{GRB SHOCKS}
\lefthead{MILOSAVLJEVI\'C \& NAKAR}

\begin{abstract}

Models for the synchrotron emission of gamma-ray burst afterglows
suggest that  the magnetic field is generated in the shock wave that
forms as relativistic ejecta plow through the circum-burst  medium.
Transverse Weibel instability efficiently generates magnetic fields
near equipartition with the post-shock energy density.  The detailed
saturated state of the instability, as seen in particle-in-cell
simulations, consists of magnetically self-pinched current
filaments.  The filaments are parallel to the direction of
propagation of the shock and are about a plasma skin depth in
radius, forming a quasi--two-dimensional structure.    We use a
rudimentary analytical model to argue that the Weibel filaments are
unstable to a kink-like mode, which destroys their
quasi--two-dimensional structure.  For wavelengths longer than than
the skin depth, the instability grows at the rate equal to the speed
of light divided by the wavelength. We calculate the transport of
collisionless test particles in the filaments experiencing the
instability and show that the particles diffuse in energy.  This
diffusion marks the beginning of thermalization in the shock
transition layer, and causes initial magnetic field decay as
particles escape from the filaments. We discuss the implications of
these results for the structure of the shock and the polarization of
the afterglow.


\end{abstract}

\section{Introduction}

Gamma-ray bursts (GRB) afterglows have been ascribed to synchrotron
emission from relativistic shocks in an electron-proton plasma.
Detailed studies of GRB spectra and light curves have shown that the
magnetic field strength in the shocked plasma (the downstream) is a
fraction of $\epsilon_B\sim 10^{-2}- 10^{-3}$  of the internal
energy, while the energy in the emitting electrons is a fraction of
$\epsilon_e\sim 10^{-1}$ of the internal energy
\citep{Panaitescu:02,Yost:03}. Recently, \citet{Eichler:05} have
shown that if only a small fraction $f$ of the electrons are
accelerated into the high-energy nonthermal tail, the observations
can be fitted with values of $\epsilon_B$ and $\epsilon_e$ that are
smaller by a factor of $f$ than the above estimates, as long as $f$
is larger than the electron to proton mass ratio $m_e/m_p$. This
still implies a magnetic field with $\epsilon_B\gtrsim 10^{-6}$.
Furthermore, the measurement of linear polarization at the level of
a few percent (\citealt{Bjornsson:04,Covino:04} and references
therein) implies that the magnetic field in the synchrotron emitting
region must deviate from isotropy.

Simple compressional hydrodynamic amplification of a pre-existing
magnetic field of the unshocked plasma (the upstream)
results in  $\epsilon_B\sim 10^{-9}$ \citep{Gruzinov:01}. Thus the
requisite magnetic field must be generated in the shock itself or
in the downstream. A leading candidate mechanism
that produces a magnetic field near equipartition
in the shock transition layer is the transverse
Weibel instability \citep{Weibel:59,Fried:59},
as was suggested by \citet{GruzinovWaxman:99}
and \citet{Medvedev:99}.\footnote{See
\citet{Bret:05} for a discussion of different modes of the
Weibel instability.}  This instability and the magnetic field it produces
 are expected to play a crucial role in the
thermalization of the upstream and in the shock dynamics.

Recently, a great progress toward understanding unmagnetized
collisionless shocks has been made by means of two- and
three-dimensional particle-in-cell (PIC) simulations
(e.g.,~\citealt{Lee:73,Gruzinov:01,Silva:03,
Frederiksen:04,Jaroschek:04,Medvedev:05,Kato:05}). Although the
simulations do not resolve the $e$--$p$ shock, they do show that
within a layer $\sim 100$ proton skin depths wide, where the upstream and the downstream plasma interpenetrate, the transverse
Weibel instability saturates.

The saturated state of the transverse Weibel instability consists of
magnetically self-pinched current filaments (see, e.g., Fig.~1 in
\citealt{Frederiksen:04}). The filaments are initially about a proton
plasma skin depth in diameter and are parallel to the direction of
shock propagation. As such they come with a magnetic field close to
equipartition ($\epsilon_B\sim 0.1$) that lies in the plane
perpendicular to the direction of shock propagation.  However, the
observed emission from GRB afterglows is expected to be produced at
a distance of over $\sim10^9$ plasma skin depths from the shock
(e.g.,~\citealt{Piran:05} and references therein). Thus, even if the
filamentary picture correctly describes the transition layer of GRB
shocks, only the late evolution of these filaments is relevant for
the observed emission.

\citet{Gruzinov:01} pointed out that there is no obvious theoretical
justification for the perseverance of the magnetic field.  Since the field forms only on a small scale---the plasma skin 
depth---one would expect that it also decays 
over the small distance comparable 
to the skin depth.  On the other hand, based on a 
quasi--two-dimensional picture
of current filaments, \citet{Medvedev:05} suggest that the
interaction between neighboring filaments may result in a magnetic
field of an ever growing coherence length, whereby $\epsilon_B$ is
saturated at a finite value many plasma skin depths in the
downstream.

There are reasons to believe, however, that the
quasi--two-dimensional picture is short-lived. For example, shock
compression cannot be achieved in the region where the filaments
make such an ordered structure \citep{Milosavljevic:05}.
Lacking detailed PIC
simulations, however, the filament evolution,
the isotropization of the upstream
particles, and the production of shock jump conditions remain poorly understood.  These processes, of course, are of immense importance for the microphysics
of GRBs and collisionless
shocks in general.

The two main processes that dictate the evolution the ordered
outcome of the transverse Weibel instability are the
magnetic self-confinement of each
filament and the interactions between neighboring filaments. Here, we
use simple analytical arguments to explore the first one, namely the
stability of a single filament. In \S~\ref{sec:equilibrium}, we
present an MHD model of such current filament. In
\S~\ref{sec:stability}, we show that the filament is unstable to a
kink-like mode, and estimate the growth rate of the mode. This instability destroys the quasi--two-dimensional
geometry of the filament and  produces a significant magnetic
field in the direction parallel to shock propagation. In
\S~\ref{sec:transport}, we study the motion of collisionless test
particles in the background of a current filament undergoing the
instability. We show that charged particles confined within the
filament diffuse in energy space. We suggest that this diffusion
isotropisizes and thermalizes the upstream particles. In
\S~\ref{sec:discussion}, we discuss implications of these processes
for the magnetic field decay and the physics of GRB afterglows.

\section{Equilibrium Weibel Filament}
\label{sec:equilibrium}

Consider a magnetostatic equilibrium  composed of an infinite cylindrical
current filament extending in the $z$ direction.  Let $r=0$ be the
center of the filament, where we use cylindrical coordinates
$(r,\theta,z)$. Let the current flow in the $+z$ direction along the
central axis. Further, assume that at some radius $r=R$, the current
drops to zero.  We refer to $R$ as the radius of the filament.
At $r>R$, there
are currents flowing in other filaments in both directions ($\pm z$),
so that the total current flowing through any given plane $z={\rm
const}$ vanishes. Since we are interested in the stability of a
single filament, we ignore the effect on the filament of 
currents external to the filament.

Particles giving rise to the current must move coherently, e.g., if
there is a pinching azimuthal magnetic field $B_\theta>0$, positive
charges move in the positive $z$-direction, and  negative charges
move in the negative $z$-direction. This is possible  if the
particles do not execute a full gyration in the magnetic field
\citep{Alfven:39}; fractional gyrations give rise to directed
current as explained in \citet{Spitzer:65}. This requirement
implies that a filament's radial structure
depends on its radius. A detailed discussion of this
structure can be found in \citet{Davidson:74} and \citet{Honda:00};
we here summarize the relevant aspects.

Consider first the regime in which the radius of the filament  $R$
is smaller than about the plasma skin depth, $R \lesssim \delta$.
The skin depth equals $\delta\equiv c\gamma^{1/2}/\omega_{\rm
p}$, where $\omega_{\rm p}\equiv (4\pi e^2 n/m)^{1/2}$ is the plasma
frequency, $\gamma$ is the kinetic Lorentz factor the particles, $m$
is the mass of the particles, and $n$ is their density
(in $e^\pm$ shocks, $m$ is the
electron mass; in $e^-p$ shocks, it is the proton mass). This
directly implies that the current flowing through the filament $I
\sim \pi R^2 ne\beta_\parallel c$, where $\beta_\parallel c$ is
the average axial velocity of particles in the filament, does not
exceed the Alfv\'en  critical current $I_{\rm A} =
\gamma\beta_\parallel m c^3/e$.  Since the magnetic field
is related to the current via $B_\theta \sim I/Rc$, the condition
$I\lesssim I_{\rm A}$ implies that the Larmor
radius of the particle $r_{\rm L}=\gamma \beta m c^2 / e B_\theta$
must exceed the radius of the filament $ r_{\rm L} \gtrsim
(\beta/\beta_\parallel) R > R$. As a result, when $R \lesssim \delta$, all particles magnetically
confined to the filament move in directed fashion.  Here
$\beta_\parallel$ depends weakly on $r$ and thus the current profile across the filament is approximately homogeneous.

However when $R > \delta$, if the current were uniform
within $r\lesssim R$ and the average axial velocity of the current
carrying charges were relativistic ($\beta_\parallel\sim 1$), the
Larmor radius would be smaller than the radius of the filament, and
directed motion of charges within the filament would be compromised.
Directed current flow can still be maintained when $R>\delta$ if the
current is confined within a thin annular cylindrical region of width
$\Delta R$ not exceeding the Larmor radius.  The current then gives
rise to a thin magnetic ``wall'' against which a particle confined
to the filament can be reflected.  Then the Larmor radius within the
current carrying layer can be written $r_{\rm L}=\delta^2/\Delta R$,
and the condition $r_{\rm L}\geq\Delta R$ implies that the current
carrying layer is thinner than the skin depth, $\Delta R\leq \delta$,
as seen in PIC simulations. In this regime, the saturated phase of
the transverse Weibel instability consists current-carrying domains
separated by thin magnetic walls.

Analytic considerations \citep{Milosavljevic:05} and numerical
simulations indicate that  once the transverse Weibel instability
saturates in collisionless shocks, the filament size is comparable
to the skin depth. Motivated by these results we focus here on the
case in which $R \sim \delta$.

We assume that MHD equations apply and that the pressure tensor of
the particles is isotropic.  Both of these assumptions are
oversimplifications. The pressure tensor near the axis of the
filament is approximately isotropic when $r_{\rm L}\sim R \sim
\delta$.  The purpose of the MHD model is to elucidate the physical
mechanisms and motivate specific collisionless PIC simulations of
the saturated state of the Weibel instability. The simulations are
the best way to test the theory over a range of parameter
values.

The toroidal magnetic field ${\bf
B}=B(r){\bf \hat \theta}$ is related to the axial current density
${\bf J}=J(r) {\bf \hat z}$ via
\beq
\label{eq:current_density}
4\pi
J = \frac{1}{r}\frac{d}{dr} (rB) .
\eeq
Within the MHD approximation, the fluid pressure $P(r)$
satisfies the equation of pressure equilibrium
\bea
\label{eq:grad_pressure}
\nabla P &=& {\bf J}\times {\bf B} \nonumber\\
&=& -\frac{1}{8\pi r^2} \frac{d}{dr}(r^2 B^2)  {\bf \hat r} .
\eea
To construct an magnetostatic equilibrium filament, one can choose
the radial dependence of the magnetic field, and then evaluate the
current density and the pressure using equations
(\ref{eq:current_density}) and (\ref{eq:grad_pressure}).

\section{Stability}
\label{sec:stability}

It immediately follows from equation (\ref{eq:grad_pressure}) that
the fluid pressure inside the filament is larger than outside, the
magnetic pressure accounting for the difference.  This pressure
imbalance  is the origin of the unstable behavior that we explore
below. The filaments are unstable to the well-known sausage, kink,
and related MHD modes (see., e.g., \citealt{Hasegawa:75} and
references therein) which result in the distortion of the filament
boundary.  We here focus on a particular mode, the helical kink
instability, but expect similar stability criteria, growth rates,
and particle transport in other related modes.

Consider linear magnetostatic perturbations around the equilibrium
described in \S~\ref{sec:equilibrium}.  According to the energy principle, elegantly proven in \citet{Kulsrud:05}, a perturbation given by the
Lagrangian displacement ${\vec \xi}$ grows if the associated
potential energy change is negative.  The potential energy change
is given by \citep{Bernstein:58} \beq \label{eq:delta_w} W =
\frac{1}{2} \int \left[\frac{Q^2}{4\pi}+{\bf J}\cdot({\bf
\xi}\times{\bf Q})+\gamma P(\nabla\cdot{\vec \xi})^2+({\bf
\xi}\cdot\nabla P)(\nabla\cdot{\vec \xi})\right] d{\bf x} \eeq where
${\bf Q}\equiv \nabla\times({\vec \xi}\times{\bf B})$ is the magnetic
field perturbation, and the volume integral extends over all space
in the radial direction and an averaging over $z$-direction is
assumed.

Following \citet{Newcomb:60}, we consider a perturbation with
displacement ${\vec \xi}=(\xi_r,\xi_\theta,\xi_z)$ of the form \bea
{\vec \xi} = {\rm Re}[(\xi_r,i\xi_\theta,i\xi_z)  e^{i(m\theta+kz)}]
\eea where $\xi_r(r)$, $\xi_\theta(r)$, and $\xi_z(r)$ are real
functions of radius.  Newcomb shows that non-axisymmetric
perturbations $(m\neq 0)$ that minimize $W$ are incompressible,
$\nabla\cdot{\vec \xi}=0$, and have form \bea
\label{eq:displacement_incompressible}
\xi_r &=& \xi  ,\nonumber\\
\xi_\theta &=& \frac{i}{m} \left[\frac{d}{dr}(r\xi) - \frac{k^2r^4}{k^2r^2+m^2}\frac{d}{dr} \left(\frac{\xi}{r}\right)\right] ,\nonumber\\
\xi_z &=& \frac{ikr^3 }{k^2r^2+m^2}\frac{d}{dr} \left(\frac{\xi}{r}\right) .
\eea
Note that the function $\xi$, identical to the radial displacement $\xi_r$, has been left unspecified.

For perturbations of the form in equation
(\ref{eq:displacement_incompressible}), the potential energy
perturbation in equation (\ref{eq:delta_w}) can be expressed in
terms of the radial displacement only, 
\beq 
W = \frac{\pi}{2}
\int_0^\infty \left[f \left(\frac{d\xi}{dr}\right)^2 + g\xi^2\right]
dr , \eeq where \bea
f&\equiv& \frac{rm^2B_\theta^2}{k^2r^2+m^2} ,\nonumber\\
g&\equiv& \frac{1}{r} \frac{m^2B_\theta^2}{k^2r^2+m^2} + \frac{1}{r}
m^2 B_\theta^2 -
2\frac{B_\theta}{r}\frac{d}{dr}(rB_\theta)\nonumber\\&
&+m^2\frac{d}{dr}\left(\frac{B_\theta^2}{k^2r^2+m^2}\right) , 
\eea
as shown in equations (16-18) of \citet{Newcomb:60} (note that in
our case $B_z=0$).

The perturbations with $m=1$ are special in that they need not
vanish at $r=0$ and do not incur substantial cost in magnetic energy
as the field lines are bent only minimally. Therefore one is allowed
to assume that the radial displacement is independent of radius,
$d\xi/dr = 0$, over the region with the non-vanishing magnetic
field, which greatly simplifies the analysis. The true fastest
growing mode may not have constant $\xi$, but one expects the
constant $\xi$ approximation to come close.  For these perturbations
the radial displacements are 
\bea 
\label{eq:xi_solid}
\xi_r &=& \xi = {\rm const}  ,\nonumber\\
\xi_\theta &=& i \left(1+\frac{k^2r^2}{k^2r^2+1}\right) \xi ,\nonumber\\
\xi_z &=& -\frac{ikr }{k^2r^2+1} \xi . 
\eea 
With this, the energy
per unit length along the $z$-direction becomes 
\beq
\label{eq:energy_solid_shift} W = -\frac{\pi}{2} k^2 \xi^2
\int_0^\infty \frac{rB_\theta^2}{k^2r^2+1}   dr . 
\eeq 
Equation (\ref{eq:energy_solid_shift}) tells us that all wave numbers $k>0$
are unstable. In the long and the short wavelength limits, the
energy can approximately be written as 
\beq
\label{eq:energy_approximate} 
W \approx \cases{-\onequarter \pi\xi^2
k^2 R^2\bar B_\theta^2  , &  ($kR\ll 1$) ,\cr -\onehalf\pi \xi^2
\tilde  B_\theta^2 , & ($kR\gg 1$) \cr } \eeq where  $\bar B_\theta$
is the RMS magnetic field, and $\tilde B_\theta$ is the magnetic RMS
field averaged per unit log-radius.

Equation (\ref{eq:energy_solid_shift}) tells us that all $m=1$ modes
with finite wavelengths ($k>0$) are unstable.  An estimate of the
linear growth rate is given by \beq \label{eq:estimate_gamma}
\Gamma\sim \sqrt{-\frac{W}{K}} , \eeq where \beq \label{eq:k}
K\equiv \frac{1}{2} \int \rho (\xi_r^2+\xi_\theta^2+\xi_z^2)
rd\theta dr . \eeq The value of $K$ for perturbations in equations
(\ref{eq:xi_solid}) equals \beq
K=\frac{\pi\bar\rho\xi^2}{4k^2}[5k^2R^2-3\ln(k^2R^2+1)] \eeq where
$\bar\rho$ is the average mass density.  In the long and the short
wavelength limit, therefore, \beq \Gamma\sim \cases { k\bar
B_\theta/\sqrt{2\bar \rho} , & ($kR\ll 1$) ,\cr
 R^{-1} \tilde B_\theta/\sqrt{5\bar \rho/2} , & ($kR\gg 1$)  .\cr
} \eeq Note that the growth rates are proportional to the
``nonrelativistic'' Alfv\'en velocity  $\bar v_{\rm A}\equiv
B_\theta / \sqrt{4\pi \bar \rho}$ multiplied by $k$ and $R^{-1}$,
respectively, in the long and the short wavelength limit. This is
expected since the instability is driven by a pressure imbalance.
The growth rate in the long wavelength limit can also be expressed
as $\Gamma\sim (\pi/2)^{1/2}\beta_\parallel kR  \omega_{\rm p}$,
where as before $\omega_{\rm p}$ is the plasma frequency.

Note that the magnetic field acquires a component parallel to the axis of the filament.  The RMS strength of the parallel field is $\bar B_z=\bar Q_z\sim \onehalf k \xi B_\theta$ in the long wavelength limit.

The above analysis is valid only as long as the perturbation is
nonrelativistic (i.e., $\xi<c/\Gamma$) and  the 
wavelength of the perturbation 
is larger than the Larmor radius.  Since we here consider the filaments with
$r_{\rm L}\sim R \sim \delta$, the analysis is valid only for 
wavelengths with $2\pi/k\gtrsim r_{\rm L}\gtrsim \delta$. 
In such filaments the
growth rate is $\Gamma \sim \beta_\parallel ck$.

\section{Transport in an Unstable Filament}
\label{sec:transport}

We next address the orbits of collisionless test particles in the electromagnetic field defined by the magnetostatic equilibrium (\S~\ref{sec:equilibrium}) and the linear unstable perturbation (\S~\ref{sec:stability}).  Consider a particle with unperturbed orbit confined to the region $r<R$, and assume that the motion is directed along the axis, $dz/dt>0$. The momentum of the particle ${\bf p}$ is then a periodic function of $z$, as are its radial excursion and azimuth.   The magnetostatic equilibrium is electrically neutral and the energy of the particle, ${\cal E}=\gamma mc^2$, where $\gamma\equiv(1+p^2/m^2c^2)^{1/2}$, subject to the magnetic field of the equilibrium only, is a constant of motion.  The time variation of the magnetic field of the unstable perturbation induces an electric field given by Ohm's law
\beq
\label{eq:electric_field_perturbation}
{\bf E}= -\frac{{\bf V}\times{\bf B}}{c} ,
\eeq
where ${\bf B}$ is the magnetic field of the equilibrium, and ${\bf V}\equiv d{\vec \xi}/dt$ is the bulk velocity associated with the perturbation.   In long wavelength perturbations, the displacement is mainly perpendicular to the $z$-axis and the electric field is thus mainly parallel to the axis.

In writing equation (\ref{eq:electric_field_perturbation}), we have assumed infinite conductivity, and have ignored terms of the form ${\bf J}\times {\bf B}$, $\partial {\bf J}/{\partial t}$, and $\nabla P$. These terms can be neglected to the first order in $k\delta$ and $\Gamma/\omega_{\rm p}$, i.e., for wavelengths longer than the plasma skin depth.

Kinetic energy of a test particle with charge $q$ incurs a change of
\beq
\Delta {\cal E} = q \int_{-\infty}^t \frac{{\bf p}\cdot {\bf E}}{\gamma m} dt' .
\eeq
To the first order in the perturbation, the energy change equals
\bea
\label{eq:energy_change_integral}
\Delta {\cal E} &=& - q\Gamma \int_{-\infty}^t {\bf \beta}\cdot ({\vec \xi}\times {\bf B}) dt' \nonumber\\
&=&-q\Gamma \int_{-\infty}^t [\xi_z\beta_r \sin(kz+\theta)  \nonumber\\ &&
\ \ \ \ \ \ \ \ \ \ \ \ \ + \xi_r\beta_z \cos(kz+\theta)] e^{\Gamma t'}B_\theta dt' ,
\eea
where $\beta\equiv{\bf p}/\gamma mc$ and as before $\Gamma$ denotes the growth rate of the instability.

The energy change can be evaluated for the incompressible perturbations given in equations (\ref{eq:xi_solid}).  The displacement of the fluid is an oscillatory function of the monotonic variable $z(t)$, as is the displacement of the particle.  Resonances occur when $\beta_\parallel ck\sim \omega_\theta$, where $\beta_\parallel c$ is the average axial velocity of the particle, and $\omega_\theta$ is its azimuthal angular frequency, which will differ from one particle to another.  At a resonance, the particle experiences a coherent electric field over most of its orbit.

\section{Thermalization}
\label{sec:thermalization}

To explore the effect of energy change derived in equation (\ref{eq:energy_change_integral}), we consider the simplest Maxwell-Vlasov equilibrium of a magnetically self-pinched current filament \citep{Hammer:70,Davidson:74}, in which all particles have the same total energy ${\cal E}=\gamma_0 mc^2$ and the same axial canonical momentum ${\cal P}_z=\beta_\parallel\gamma_0 mc$, both of which are constants of motion. Here, $\gamma_0$ is the Lorentz factor of the particles and $0<\beta_\parallel<(1-\gamma_0^{-2})^{1/2}$ is a constant. The canonical momentum equals ${\bf \cal P}={\bf p}-(e/c){\bf A}$, where ${\bf A}$ is the vector potential.

Particle density $n$ inside the filament associated with this equilibrium is uniform.
The equilibrium has phase-space distribution function
\beq
\label{eq:distribution}
f({\cal E},{\cal P}_z)=\frac{n}{2\pi\gamma_0 m}\delta({\cal E}-\gamma_0 mc^2)\delta({\cal P}_z-\beta_\parallel\gamma_0 mc) ,
\eeq
where $\delta(x)$ is the Dirac delta-function, not to be confused with the skin depth. The magnetic field and the vector potential inside the filament are given by
\bea
\label{eq:model_field}
B_\theta&=&-\frac{\beta_\parallel\gamma_0 mc^2}{e\delta} I_1\left(\frac{r}{\delta}\right) , \nonumber\\
A_z&=&-\frac{\beta_\parallel\gamma_0 mc^2}{e} \left[1-I_0\left(\frac{r}{\delta}\right)\right] ,
\eea
where as before $\delta$ is the plasma skin depth inside the filament, and $I_n(x)$ is the modified Bessel function of the first kind.  The filament radius $R$ is the solution of the equation
\beq
I_0^2\left(\frac{R}{\delta}\right)=\frac{\gamma_0^2-1}{\gamma_0^2\beta_\parallel^2} .
\eeq
Although extremely simple, this model correctly captures many of the characteristics of current filaments seen in PIC simulations, including the concentration of the current flow near the edge of the filament when $R\gtrsim\delta$.

\begin{figure}
\epsscale{1.2}
\plotone{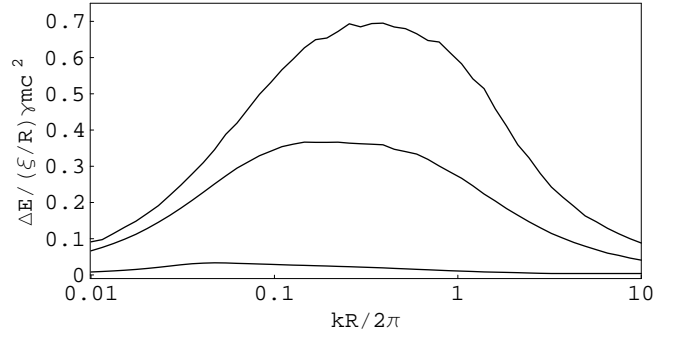}
\caption{The RMS linear energy change $\Delta{\cal E}$ of a particle in units of $(\xi/R) \gamma mc^2$ as a function of the wave number $k$ for the model described in \S~\ref{sec:thermalization} with $\beta_\parallel=(0.05,0.5,0.95)$ from top to bottom and $\gamma_0=10$.  Here, $\xi$ is the final value of the displacement associated with the instability discussed in \S~\ref{sec:stability}. The energy change was calculated using Monte-Carlo integration. \label{fig:delta_e}}
\end{figure}

Figure \ref{fig:delta_e} shows the RMS linear energy change of particles drawn from the distribution of equation (\ref{eq:distribution}) with $\beta_\parallel=(0.05,0.5,0.95)$ as a function of the wave number $k$ of the unstable perturbation.  The magnetic field of the equilibrium filament is given by equations (\ref{eq:model_field}), and the filament is undergoing instability with displacement given in equations (\ref{eq:xi_solid}) and growth rate defined by equations (\ref{eq:energy_solid_shift}), (\ref{eq:estimate_gamma}), and (\ref{eq:k}).  For $\beta_\parallel\gtrsim 0.5$, the linear energy change is maximum for $kR\sim (0.1-0.4)\times2\pi$.  It should be kept in mind, however, that the pressure tensor in filaments with $\beta_\parallel\ll 0.5$ and those with $1-\beta_\parallel\ll 0.5$ is anisotropic, and thus the growth rates calculated in \S~\ref{sec:stability} may not be accurate.

The instability affects in a similar way the pitch angle of the
particles; the pitch angle can be defined as the angle subtended to
the central axis at the instance of closest approach to the axis. We
have focused on the energy because it is a simpler calculation.

As the orbits of particles inside a filament are perturbed by the instability, some of the particles that are initially confined to the filament  escape.  For example, consider a particle moving in the meridional plane ($\theta={\rm const}$) that has radial velocity $\beta_r$ when crossing the central axis.  Assume that the particle is marginally confined to the filament, i.e., $\beta_r^2 \sim R/r_{\rm L}$, where $r_{\rm L}\sim \gamma m c^2 / e\bar B_\theta$ is the approximate Larmor radius.  Evidently, an increase of the pitch angle  $\phi=\sin^{-1}[\beta_r(1-\gamma^{-2})^{-1/2}]$ at a constant $\gamma$, or an increase of the energy ${\cal E}=\gamma mc^2$ at a constant pitch angle, both caused by the perturbed electromagnetic field of the instability discussed above, can liberate the particle from the filament.

A filament experiencing a growing kink instability continuously sheds particles in this fashion. The liberated particles have sufficient energy to visit neighboring filaments of either sign of the current, and can be thought of having joined a thermalized pool.  The loss of particles implies decrease of the current flowing through the filament, which in turn implies the decay of the magnetic field.  Only the particles that remain confined to the filaments contribute to the coherent toroidal magnetic field of the filament. Rapid field decay can be partially offset by an increase in the current per particle by electric fields induced during flux loss; robustness of the currents depends on the competition between the scattering and the induction.

\section{Discussion}
\label{sec:discussion}

We have argued that the quasi--two-dimensional structure of the
transition layer in collisionless shocks is disrupted by
pressure-driven instabilities.  We have also shown that the
disruption is accompanied by a redistribution of particle energies,
which can be interpreted as the onset of thermalization and magnetic
field decay.  Long term evolution of the magnetic field must
therefore be addressed in the context of the three-dimensional
turbulence that ensues after the Weibel filaments have been
disrupted. The latter regime remains poorly understood.

An alternate interpretation of the early decay of the magnetic field
refers to the hierarchical merging of current filaments due to
Lorentz forces \citep{Gruzinov:01,Medvedev:05,Kato:05}. It can be
asked whether the merging or the pressure-driven instabilities  prevail.  Since
the maximum current that can flow through a filament is limited (see
\S~\ref{sec:equilibrium}), magnetic energy density must eventually
decay under merging.  However, PIC simulations of $e^\pm$ shocks
(Spitkovsky \& Arons 2005, private communication) show that the
current inside a filament is tightly shielded by a reverse current
flowing just outside the filament, i.e., opposite current filaments
are tightly packed and most of the current flows near the edge. The
shielding currents reduce the Lorentz attraction  and slow the
growth of the magnetic field correlation length via merging.

Meanwhile, the filaments are susceptible to the instabilities
independently of the shielding.  Therefore, we expect the dynamics
of single filaments to be governed by the instabilities. In particular, the instabilities may drive merging between the filament fragments in three dimensions.
The highest-resolution published simulations of
cold shell collisions with two particle species by
\citet{Frederiksen:04} show clear evidence for progressive bending
and kinking of the ``proton'' ($m_p/m_e=16$) filaments (see their
Fig.~2).

Similar mechanism was recently studied by \citet{Zenitani:05}, who
carried out two-dimensional PIC simulations of the relativistic
drift-kink instability in an infinite $e^\pm$ current sheet confined
between  reversed magnetic fields.  As the sheet bends in the
direction perpendicular to the magnetic field, an alternating
electric field is induced parallel to the sheet.  The authors compare
the growth rates measured in the simulations with predictions from
two-fluid theory and find good agreement for $k\lambda\lesssim 0.7$,
where $k$ is the wave number and $\lambda$ is the thickness of the
current sheet. They also find that in the nonlinear stage of the
instability, the electric field becomes coherent in the central region of
the current sheet, which accelerates particles and efficiently
dissipates the magnetic energy energy.

If external shocks in GRBs resemble the observed outcome of
relativistic shell collisions in PIC simulations, the instability
discussed here has implications for the interpretation of the
observed weak linear polarization of the afterglow. The optical
emission of GRB afterglows is linearly  polarized at a level of a
few percent, implying that the magnetic field in the emitting region
is anisotropic. There are two different forms of magnetic field
anisotropy that can produce this polarization. The first is a  field
parallel to the plane of the shock that is coherent on scales
exceeding the plasma skin depth by many orders of magnitude
\citep{GruzinovWaxman:99,Granot:03}. The second is a combination of
a random magnetic field within the plane of the shock and a
non-axisymmetric geometry of the emitting region
\citep{Gruzinov:99,Sari:99,Ghisellini:99,GranotKonigl:03,Nakar:03,Rossi:04}.

A coherent magnetic field within the plane of the shock is unlikely
to be produced in the shock itself since there is no preferred
direction within the plane of the shock. Such field could in principle result by amplification
of a pre-existing, ambient coherent magnetic field.  This possibility, however,
is implausible because the field in the
 medium into which the GRB ejecta plow is expected to be far too weak. The
second possibility is that the coherence length of the magnetic field 
generated the shock grows rapidly. Since at any given time the
observer sees many causally disconnected regions
\citep{GruzinovWaxman:99,Nakar:04}, the coherence length of the
downstream magnetic field should grow  at a rate close
to the speed of light in order to produce a
polarization at the level of a few percent. 
We do not know of a mechanism that would facilitate
such growth, especially given that the Alfv\'en speed in the weakly
magnetized relativistic downstream plasma is much smaller than the
speed of light.

An anisotropic field, but random within the plane of the shock,
appears to be a much more appealing possibility because the shock
breaks the isotropy and could in principle result in a different
mean field strength in the parallel and the perpendicular direction.
In this case the level of polarization also depends on the degree to
which the axisymmetry of the emitting region is broken. Assuming
that the magnetic field lies entirely in the plane of the shock
(i.e., $B_z=0$), and that the geometry of the emitting region is as
expected during the jet-break in the light curve, the level of
polarization is at most $20\%-30\%$ \citep{Sari:99,Rossi:04} and in
some scenarios it can be as small as a few percent. If $B_z \approx
B_{xy}$,  the polarization level is significantly reduced.
Therefore, the typical observed polarization of a few percent requires that
either $B_z < B_{xy}/2$ or $B_z > 2B_{xy}$. Even this  minor
difference between $B_z$ and $B_{xy}$ presents a theoretical
challenge in view of our results, which suggest that $B_z$ arises
quickly within the transition layer as the current filaments associated  with the small-scale magnetic fields are effectively destroyed.
We speculate that the magnetic field quickly becomes isotropic in
the rest frame of the shocked plasma.

\acknowledgements

We are indebted to P.\ Goldreich and T.\ Piran for comments and advice, and J.\ Arons and A.\ Spitkovsky for many inspiring discussions and for sharing their numerical simulations with us ahead of publication.  This work was supported at Caltech by a postdoctoral fellowship to M.~M.\ and a senior research fellowship to E.~N.\ from the Sherman Fairchild Foundation.

\end{document}